# Variable Focus Nonlinear Lens via Transformation Optics

**Apra Pandey and Natalia M. Litchinitser**

*Electrical Engineering Department*

*University at Buffalo, The State University of New York*

*230L Davis Hall, Buffalo, NY 14260*

apandey4@buffalo.edu

**Abstract:** We propose a structure consisting of nonlinear core and graded-index shell that can be switched from a concentrator squeezing light into the core to a variable focus lens by varying the intensity of incident light.

**OCIS codes:** (160.3918) Metamaterials; (130.4310) Nonlinear; (130.3120) Integrated optics devices

## 1. Introduction

By leveraging the capabilities of photonics (speed) and of electronics (compactness), it should be possible to realize high performance integrated opto-plasmonic systems with applications from high bandwidth communications to sensing, and beyond. Such integration requires the availability of ultra-compact, ultra-fast, reconfigurable and tunable photonic components.

In this work, we propose and design one of such reconfigurable electromagnetic (EM) components with variable focus such that its output field profile can be tuned from an unfocussed beam to a highly localized beam with extended focal region that can be moved from infinity towards the lens surface by changing the intensity of the incident beam. Two configurations of such device will be discussed: i) self-focusing beam and ii) all-optically externally controlled focusing. The proposed device is designed using transformation optics [1,2] and in particular, based on concentrator reported by Rahm et al. [3] and Wang et al. [4].

## 2. Nonlinear lens theory and modeling

In cylindrical coordinate system, the space of a concentrator is divided into two regions; the core and the shell. The shell of the concentrator channels the incident field into the core where it is concentrated. The transformation needed to 'squeeze' light into the core in parameters can be described by [3, 4]:

$$r' = \begin{cases} R_1 r / R_2 & 0 \leq r < R_1 \\ (R_3 - R_1) r / (R_3 - R_2) - (R_2 - R_1) R_3 / (R_3 - R_2) & R_1 \leq r \leq R_3 \end{cases} \quad (1)$$

This transformation is mapped to material tensors as follows [3, 4]:

$$\varepsilon_r^{i,j} = \mu_r^{i,j} = \begin{cases} \begin{bmatrix} 1 & 0 & 0 \\ 0 & 1 & 0 \\ 0 & 0 & (R_2/R_1)^2 \end{bmatrix} & 0 \leq r \leq R_1 \\ \begin{bmatrix} (r+M)/r & 0 & 0 \\ 0 & r/(r+M) & 0 \\ 0 & 0 & N(r+M)/r \end{bmatrix} & R_1 < r \leq R_3 \end{cases} \quad (2)$$

where $M = (R_2 - R_1)R_3/(R_3 - R_2), N = [(R_3 - R_2)/(R_3 - R_1)]^2$

Finite element method is used to simulate the working of the designed tunable lens. The schematic of the proposed device designed to function at a wavelength of 1.5 µm is as shown in Fig. 1a. Dimensions of the structure are such that $R_1$ = 2µm, $R_2$= 4µm and $R_3$= 6µm. Computation domain is terminated by perfectly matched layers (PMLs) to absorb the scattered field. Structure is illuminated by a TE polarized plane wave incident from left having a wavelength of 1.5 µm. Graded refractive index of the shell region squeezes this light towards the core.

When the applied incident field is low, the device acts in a linear regime and behaves like the concentrator as was demonstrated in [3, 4] and concentrates the EM field inside the core region. The surface plot of power flow for this case is as shown in Fig. 1b.

Tunable lens proposed here utilizes a nonlinear core element. We use Kerr-type nonlinear medium having a cubic nonlinearity defined by $\chi^{(3)}$. Nonlinear phenomena change the optical properties of the core as the refractive index of the core dependent on the intensity of the incident electric field. On increasing the intensity of the incident field, the refractive index of the core increases. Electromagnetic radiation is slowed down inside the core and the part of the wave that travels the farthest is delayed the most, which results in a net curvature and hence focusing of the field. The device operates as a lens and focusses field such that the intensity is confined to a localized narrow spot. The location of that focal spot moves from infinity (linear case) towards the lens as shown in Fig. 2 (a)-(c). Upon further increasing the applied field, the refractive index of the core becomes sufficiently high such that the device becomes a reflector/scatter and reflects away all the EM radiation.

## 3. Conclusion

We demonstrate a tunable structure that can be switched from the functionality of a concentrator to that of a lens and to a scatter/reflector depending on the intensity of the incident light. This device may find applications for the development of ultra-compact reconfigurable optical components for all optical circuits.

This research was supported by the US Army Research Office Award # W911NF-09-1-0075.

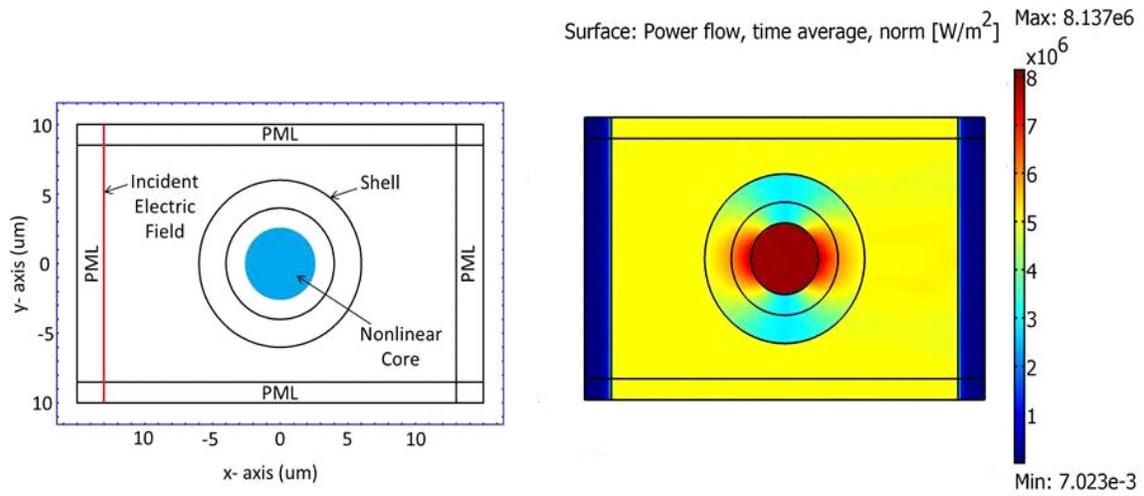

Fig. 1. (a) Schematic diagram of concentrator used in ref 3, 4 consisting of shell and a core region. Core region has a radius $R_1$. Region around the core (i.e. the shell) extends up to $R_3$. (b) Power flow distribution in a device in the linear regime working like a concentrator and 'squeezing' light inside the core.

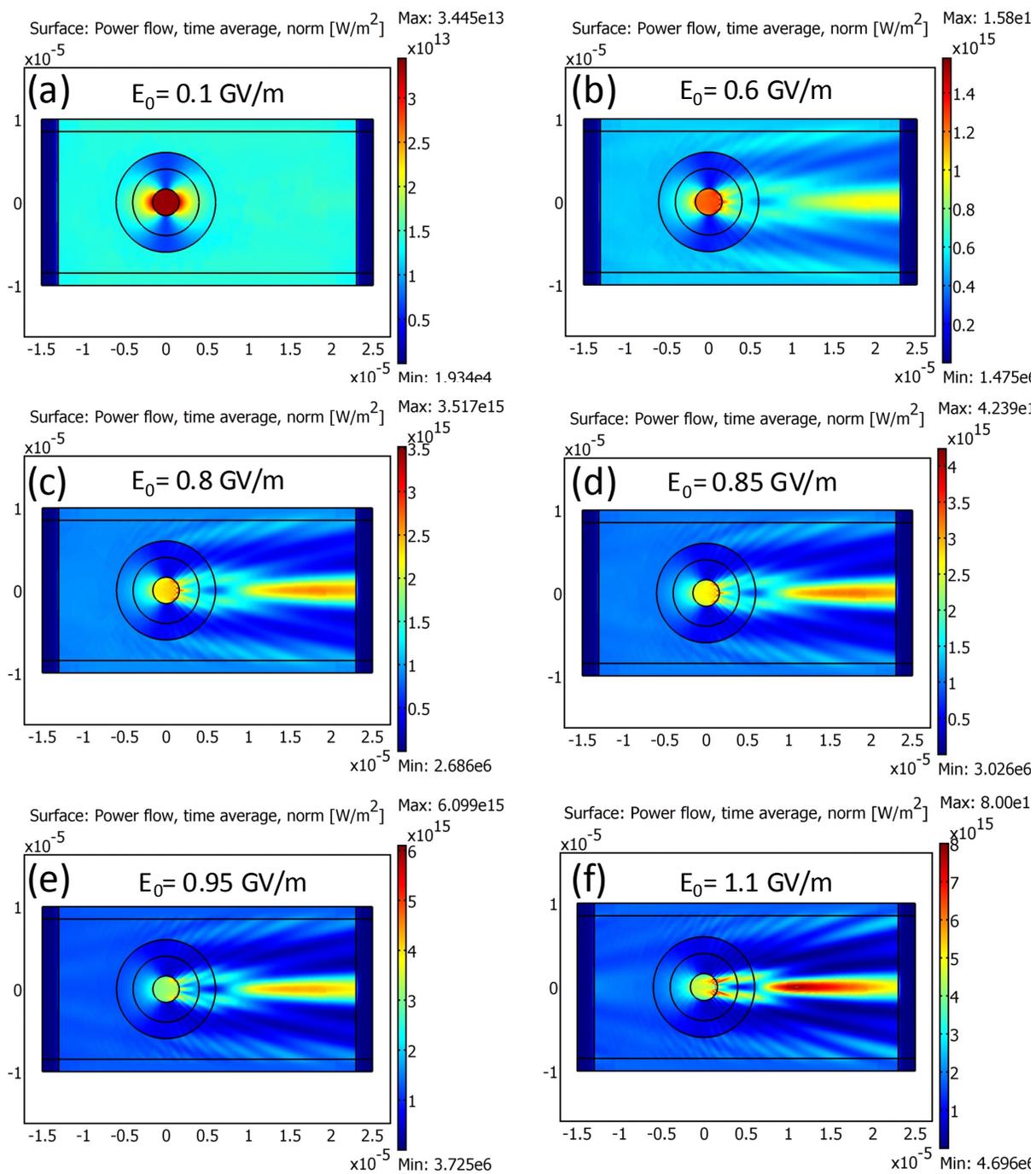

Fig. 2. Finite-element method based simulations of variable focus nonlinear lens for different values of input light intensity.